\documentclass[journal,10pt]{IEEEtran}

\usepackage{graphicx, amsmath, amssymb, lettrine, cuted, cite, xcolor}
\usepackage[justification=centering]{subfig}
\usepackage[font=scriptsize]{caption}
\DeclareMathOperator*{\argmin}{argmin}
\DeclareMathOperator*{\mymin}{min}

\begin{document}
	\title{Index Modulation Based Coordinate Interleaved Orthogonal Design for Secure Communications}
	
	\author{Burak~Ozpoyraz,~\IEEEmembership{Student Member,~IEEE}, Ibrahim~Yildirim,~\IEEEmembership{Student Member,~IEEE} and \\ Ertugrul~Basar,~\IEEEmembership{Senior Member,~IEEE} \vspace{-5.5ex}
	
	\thanks{Copyright (c) 2015 IEEE. Personal use of this material is permitted. However, permission to use this material for any other purposes must be obtained from the IEEE by sending a request to pubs-permissions@ieee.org.}
	\thanks{This work was supported by TUBITAK under Grant 218E035.}
	\thanks{The authors are with the Communications Research and Innovation Laboratory (CoreLab), Department of Electrical and Electronics Engineering, Koç University, Sariyer 34450, Istanbul, Turkey (e-mail: bozpoyraz20@ku.edu.tr; ebasar@ku.edu.tr).}
	\thanks{I. Yildirim is also with the Faculty of Electrical and Electronics Engineering, Istanbul Technical University, Istanbul 34469, Turkey (e-mail:yildirimb@itu.edu.tr).}
	\thanks{MATLAB codes, simulation results and figures are available at https://github.com/burakozpoyraz/CIOD-IM}
	}
	
	\maketitle
	\begin{abstract}
		In this paper, we propose a physical layer security scheme that exploits a novel index modulation (IM) technique for coordinate interleaved orthogonal designs (CIOD). Utilizing the diversity gain of CIOD transmission, the proposed scheme, named CIOD-IM, provides an improved spectral efficiency by means of IM. In order to provide a satisfactory secrecy rate, we design a particular artificial noise matrix, which does not affect the performance of the legitimate receiver, while deteriorating the performance of the eavesdropper. We derive expressions of the ergodic secrecy rate and the theoretical bit error rate upper bound. In addition, we analyze the case of imperfect channel estimation by taking practical concerns into consideration. It is shown via computer simulations that the proposed scheme outperforms the existing IM-based schemes and might be a candidate for future secure communication systems. 
	\end{abstract}


	\begin{figure*}[t]
		\centering
		\includegraphics[width=75ex]{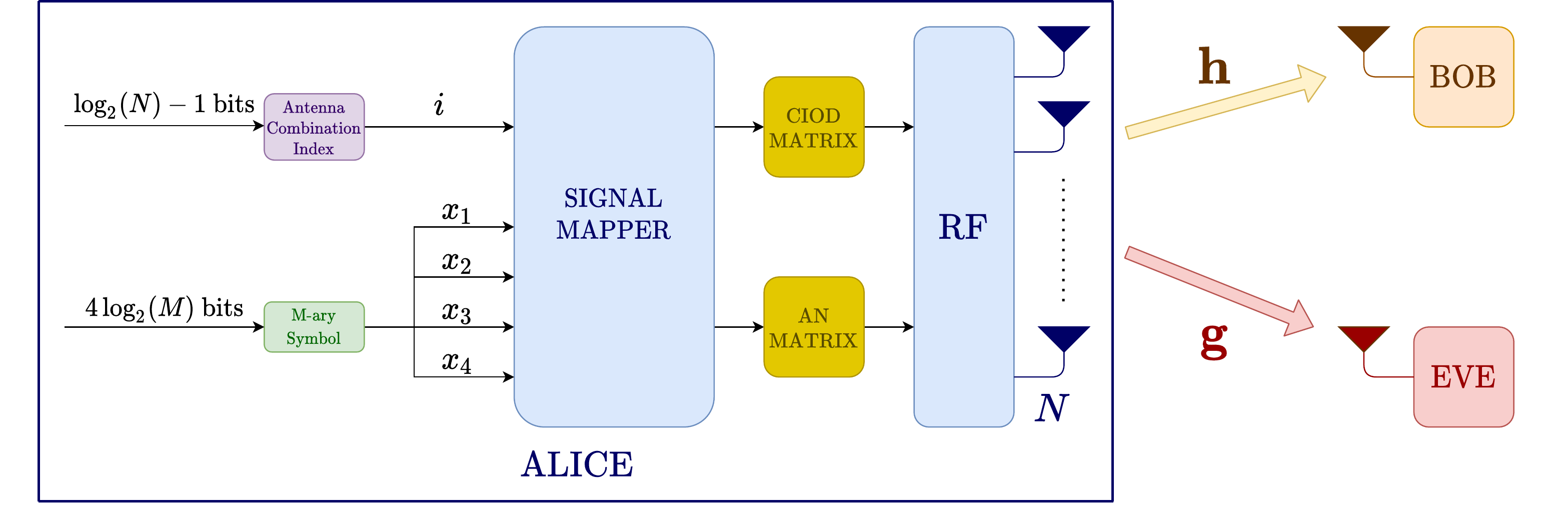}
		\caption{The system model of the CIOD-IM scheme.}
		\label{fig:SystemModel}
		\vspace{-3ex}
	\end{figure*}
	
	\section{Introduction}
	\lettrine[nindent=0em, findent=0.1em]{I}{ndex} modulation (IM) schemes have been gaining a tremendous interest over the years considering innovative ways to convey additional information \cite{1_2017_EBsurvey}. Spatial modulation (SM), which employs IM by utilizing the indices of transmit antennas as an additional information source, has been a promising advancement on conventional multiple-input multiple-output (MIMO) transmission \cite{1_2017_EBsurvey}, \cite{2_2008_SM}. In SM systems, inter-channel interference (ICI) and inter-antenna synchronization (IAS) can be prevented due to the activation of a single antenna during transmission. Besides, SM can reduce the hardware cost by using only a single RF chain. However, the broadcast nature of wireless communication channels introduces SM the risk of information leakage to eavesdroppers. Therefore, physical layer (PHY) security has attracted the attention of researchers in recent years. PHY security exploits the channel between the transmitter and the intended receiver in order to prevent information leakage. SM and PHY security exploit the same principle as they both utilize the randomness and uniqueness of wireless channels. SM based solutions originate new degrees of freedom for PHY security. There are numerous systems in the literature utilizing SM for PHY security. In \cite{3_2016_Precoding}, the authors proposed a precoding-aided SM (PSM) scheme, where the transmitter (Alice) is aware of the channel state information (CSI) of both intended receiver (Bob) and eavesdropper (Eve). Considering the passive eavesdropper case, where Alice is not aware of Eve, a secret PSM (SPSM) scheme is introduced in \cite{4_2015_SPSM}. Here, along with a precoding matrix, which is constructed by zero-forcing method, a fast time-varying precoder is proposed to provide a further enhancement in the secrecy rate. The SPSM principle is integrated to multi-user (MU) systems in \cite{5_2016_SecureMUMIMO}, where MU interference cancellation is implemented to improve the signal-to-interference plus noise ratio (SINR) at Bob. PHY security is explored using an artificial noise (AN) signal in \cite{6_2015_SecrecyEnhancement}. In this study, a jamming signal, which lies in the null space of the channel between Alice and Bob, is transmitted along with an amplitude-phase modulation (APM) symbol. Moreover, the security performance of AN-aided SM is analyzed under the case of imperfect channel estimation in \cite{7_2018_ImperfectCSI}. As a further improvement on earlier AN schemes, an AN cancellation scheme is introduced where transmit antenna selection (TAS) is applied to select a number of transmit antennas in \cite{8_2019_Conventional}. Here, two AN signals cancel out each other at Bob using the knowledge of Bob's CSI. However, the BER performance of this system model is open to improvement through diversity techniques. Thus, in \cite{9_2019_EfficientAlamouti}, Alamouti's space time block code (STBC) is utilized to provide diversity gain. Two AN signals are transmitted from the active antennas of Alamouti's scheme and they are cancelled out at Bob. Nonetheless, \cite{9_2019_EfficientAlamouti} has a strict limitation that Alice is always equipped with three antennas, which is not suitable for flexible large-scale MIMO systems. 
	
	Against this background, we propose a novel PHY security scheme for single-user (SU) multiple-input single-output (MISO) systems, which is called as CIOD-IM. The main motivation of our system is to ensure PHY security along with high spectral efficiency and diversity by means of a novel IM method and coordinate interleaved orthogonal designs (CIOD). In this system, we introduce a special AN matrix design that enables the cancellation of AN signals at Bob for each time slot. We analyzed the bit error rate (BER) as well as ergodic secrecy rate (ESR) performance over Rayleigh fading channels. In addition, we analyze the BER and ESR performance in the presence of imperfect channel estimation since it might be challenging to obtain perfect CSI in practice. Furthermore, the BER results are verified by the theoretical upper bounds. Our extensive computer simulations reveal that the proposed scheme outperforms the conventional \cite{8_2019_Conventional} and the efficient Alamouti \cite{9_2019_EfficientAlamouti} schemes in terms of the secrecy and the BER performance. Finally, we note that the CIOD-IM scheme is suitable for future low-complexity and massive MIMO systems by providing satisfactory BER results.
	\vspace{-2.3ex}	
	\section{System Model}
	In this section, we introduce the essentials of CIODs and the working principle of the CIOD-IM scheme. 
	\vspace{-2ex}		
	\subsection{Coordinate Interleaved Orthogonal Designs}
	We utilize CIOD in our system, which is a promising concept for MIMO systems. The main idea is the transmission of the in-phase and the quadrature components of APM symbols via different transmit antennas during different time slots to provide a diversity gain. Considering the specific CIOD for four transmit antennas, we note that two RF chains are required at each time slot which reduces the complexity. Conventional size 4 and rate 1 CIOD matrix is given by
	
	{\footnotesize
	\begin{align}
		\begin{split}
			\textbf{S}(x_{1},\dots,x_{4}) = 
			\begin{bmatrix}
				\boldsymbol{\theta}(\tilde{x}_{1},\tilde{x}_{2}) & \textbf{0} \\
				\textbf{0} & \boldsymbol{\theta}(\tilde{x}_{3},\tilde{x}_{4})
			\end{bmatrix} = 
			\begin{bmatrix}
				\tilde{x}_{1} & -\tilde{x}_{2}^{*} & 0 & 0 \\
				\tilde{x}_{2} & \tilde{x}_{1}^{*} & 0 & 0 \\
				0 & 0 & \tilde{x}_{3} & -\tilde{x}_{4}^{*} \\
				0 & 0 & \tilde{x}_{4} & \tilde{x}_{3}^{*}
			\end{bmatrix}
			\label{eq1}
		\end{split}
	\end{align}
	}where $\tilde{x}_{i} = \Re\{x_{i}\} + \j \Im\{x_{a}\}, i \in \{1,\dots,4\}$ with $x_{i}$ representing the $i^{\mathrm{th}}$ symbol transmitted by the CIOD matrix, $a$ equals to $3, 4, 1, 2$ for increasing values of $i$, respectively, and $(.)^{*}$ denotes the complex conjugate. It should be noted that full diversity is achieved if and only if the coordinate product distance (CPD) of the constellation set $\Omega$ is different than zero. The CPD is given by $\Lambda = \mymin_{x_{k} \neq x_{k}^{'} \in \Omega} |x_{kI}-x_{kI}^{'}|.|x_{kQ}-x_{kQ}^{'}|$ where $x_{kI}$ and $x_{kQ}$ represent the in-phase and the quadrature components of the $k^{\mathrm{th}}$ symbol, respectively. In order to satisfy this condition, constellations with $\Lambda =0$ such as QAM are rotated with a certain angle. The CPD is maximized for square lattice constellations and QPSK when the rotation angle is $\theta=31.7175^{\circ}$ and $\theta=13.2885^{\circ}$, respectively \cite{10_2004_RectangularCIOD}. Another important feature of the CIOD transmission is its symbol-by-symbol detection capability which reduces the decoding complexity \cite{11_2006_OnSTBCCIOD}.
	\vspace{-2ex}
	\subsection{Proposed CIOD-IM Method}
	We consider a large-scale MISO secure communication system with a transmitter called Alice equipped with \textit{N} transmit antennas, a legitimate receiver called Bob, and an eavesdropper called Eve, both equipped with a single receive antenna, as shown in Fig. 1. Here, $4\log_{2}(M)$ bits determine four $M$-ary APM symbols transmitted during four different time slots while $\log_{2}(N)-1$ bits determine an antenna combination from $N/2$ possible combinations. Therefore, the spectral efficiency is $l = (4\log_{2}(M) + \log_{2}(N)-1) / 4$ bits per channel use (bpcu).

	In practice, imperfect CSI can be faced due to the channel estimation errors. Therefore, we assume that the imperfect CSI of the main channel (Alice-to-Bob) is available at Alice and Bob, whereas the imperfect CSI of the eavesdropping channel (Alice-to-Eve) is only available at Eve. The main and the eavesdropping channels are given by \cite{7_2018_ImperfectCSI}
	
	{\footnotesize
	\begin{align}
		\begin{split}
			\textbf{h} &= \sqrt{1 - \sigma^{2}} \textbf{h}_{est} +
					  	  \sqrt{\sigma^{2}} \textbf{h}_{err}, \\
		    \textbf{g} &= \sqrt{1 - \sigma^{2}} \textbf{g}_{est} +
		    			  \sqrt{\sigma^{2}} \textbf{g}_{err}
			\label{eq2}
		\end{split}
	\end{align}
	}where $\textbf{h}_{est}$, $\textbf{h}_{err}$, $\textbf{g}_{est}$, and $\textbf{g}_{err}$ are $1 \times N$ vectors indicating estimates and estimation errors of the main and the eavesdropping channels, respectively, and $\sigma^{2}$ is the power of estimation error. The entries of both main and eavesdropping channels are independent and identically distributed (i.i.d) complex Gaussian variables with zero-mean and unit-variance.
	
	\subsubsection{Novel Index Modulation}
	As shown in Fig. 2, antenna combination index bits select one of $N/2$ antenna combinations, where the IM is employed. Regardless of the selected antenna combination, the box on the left is filled with $\boldsymbol{\theta}(\tilde{x}_{1},\tilde{x}_{2})$ and on the right with $\boldsymbol{\theta}(\tilde{x}_{3},\tilde{x}_{4})$. When $i^{\mathrm{th}}$ antenna combination is selected where $i \in \{0,\dots,N/2-1\}$, $(2i+1)^{\mathrm{th}}$ and $(2i+2)^{\mathrm{th}}$ antennas are active during the first two time slots while $(N-2i-1)^{\mathrm{th}}$ and $(N-2i)^{\mathrm{th}}$ antennas are active during the last two time slots, and the CIOD matrix is constructed by $\textbf{S}_{i} = \sum_{k = 1}^{4} \textbf{A}_{2k-1,i} x_{kI} + \textbf{A}_{2k,i} x_{kQ}$ where $\textbf{A}_{u,i}, u \in \{1,\dots,8\}$ are the complex weight matrices of the CIOD matrix when $i^{\mathrm{th}}$ antenna combination is selected, and $\mathbb{E}[|x_{k}|^{2}]=\alpha P_{tot}/8$, where $\alpha$ represents the power ratio allocated to the CIOD matrix, and $P_{tot}$ represents the total transmit power. For example, assuming $N=8$ and antenna combination index bits of [0 1], the transmitted CIOD matrix is obtained as\vspace{-2ex}
	
	{\footnotesize	
	\begin{align}
		\begin{split}
			\textbf{S}_{1} =
			\begin{bmatrix}
				\textbf{0} & \textbf{0} \\
				\boldsymbol{\theta}(\tilde{x}_{1},\tilde{x}_{2}) & \textbf{0} \\
				\textbf{0} & \boldsymbol{\theta}(\tilde{x}_{3},\tilde{x}_{4}) \\
				\textbf{0} & \textbf{0}
			\end{bmatrix}
			\label{eq3}
		\end{split}
	\end{align}
	}where \textbf{0} is a $2 \times 2$ all zeros matrix.
	
	\begin{figure}[h]
		\centering
		\includegraphics[width=30ex]{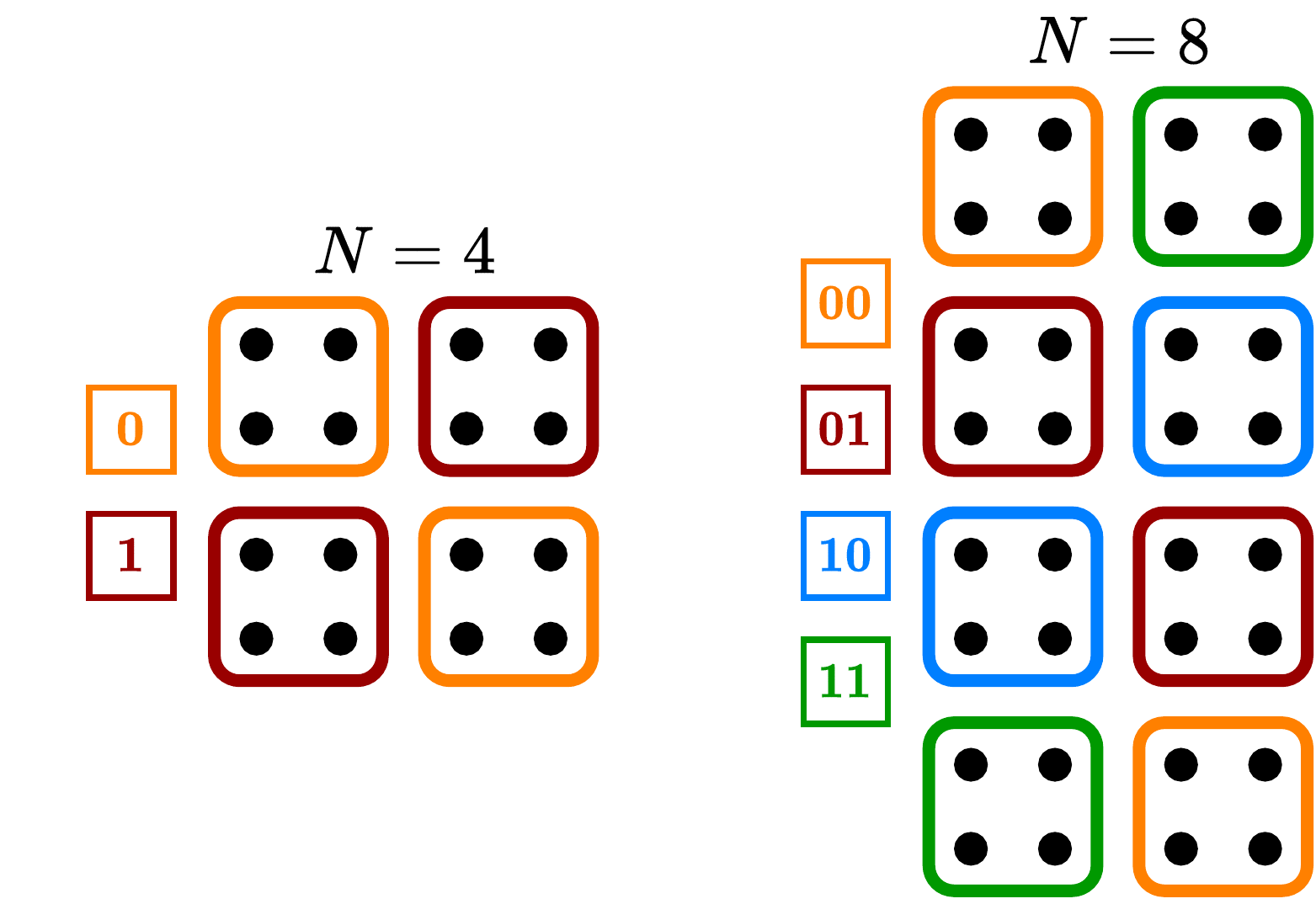}
		\caption{The antenna combinations of the CIOD-IM scheme for $N=4$ and $8$.}
		\label{fig:AntennaCombination}
		\vspace{-3ex}
	\end{figure}

	\subsubsection{Artificial Noise Cancellation}
	In order to implement AN cancellation, the AN matrix of the $i^{\mathrm{th}}$ combination, $\textbf{Z}_{i}$, should be carefully designed with respect to IM so that it is cancelled out while passing through the main channel. $\textbf{Z}_{i}$ is constructed by replacing $\boldsymbol{\theta}(\tilde{x}_{1},\tilde{x}_{2})$ and $\boldsymbol{\theta}(\tilde{x}_{3},\tilde{x}_{4})$ in $\textbf{S}_{i}$ with $\textbf{Q}(z_{11},z_{21})$ and $\textbf{Q}(z_{12},z_{22})$, respectively. Here, $\textbf{Q}$ matrices are given as
	
	{\footnotesize		
	\begin{align}
		\begin{split}
			\begin{tabular}{c c}
				$\textbf{Q}(z_{11},z_{21}) = 
				\begin{bmatrix}
					z_{11} & z_{11} \\
					z_{21} & z_{21}
				\end{bmatrix}$ 
				&
				$\textbf{Q}(z_{12},z_{22}) = 
				\begin{bmatrix}
				z_{12} & z_{12} \\
				z_{22} & z_{22}
				\end{bmatrix}$
			\end{tabular}		
			\label{eq4}
		\end{split}
	\end{align}
	}where $z_{lt} = \beta_{lt}v, l,t \in \{1,2\}$ is the AN signal transmitted from $l^{\mathrm{th}}$ active transmit antenna during $t^{\mathrm{th}}$ half of the time slots, $\beta_{lt}$ is the coefficient of the corresponding AN signal, and $v$ is a complex Gaussian random variable with zero-mean and unit-variance. When the $i^{\mathrm{th}}$ antenna combination is selected, the interference caused by $\textbf{Z}_{i}$ at Bob, $\textbf{h} \textbf{Z}_{i}$, is obtained as
	
	{\footnotesize	
	\begin{align}
		\begin{split}
			\begin{bmatrix}
				\begin{bmatrix}
					h_{2i+1} & h_{2i+2}
				\end{bmatrix} \textbf{Q}(z_{11},z_{21})
				&
				\begin{bmatrix}
					h_{N-2i-1} & h_{N-2i}
				\end{bmatrix} \textbf{Q}(z_{12},z_{22})
			\end{bmatrix}
			\label{eq5}
		\end{split}
	\end{align}
	}where $h_{j}, j \in {1,\dots,N}$ is the channel coefficient from $j^{\mathrm{th}}$ transmit antenna to Bob. When this interference equals to zero, AN signals are cancelled out at Bob. Thus, Bob is not affected by the AN matrix. However, since Alice has the imperfect CSI of the main channel, the AN coefficients are also obtained erroneously as $\beta_{11} = -h_{est}^{2i+2},\beta_{21} = h_{est}^{2i + 1}$ or $\beta_{11} = h_{est}^{2i+2},\beta_{21} = -h_{est}^{2i + 1}$, while $\beta_{12} = -h_{est}^{N-2i},\beta_{22} = h_{est}^{N-2i-1}$ or $\beta_{12} = h_{est}^{N-2i},\beta_{22} = -h_{est}^{N-2i-1}$, where $h_{est}^{j}$ is the estimate of the channel coefficient from $j^{\mathrm{th}}$ transmit antenna to Bob. It should be noted that the AN effect cannot be eliminated by Eve even with the knowledge of Alice-Bob channel because of the random effect of $v$.
	
	\subsubsection{Received Signals}
	The vector of received signals at Bob is given as
	
	{\footnotesize
	\begin{align}
		\begin{split}
			\textbf{y}_{B} &= \textbf{ } \textbf{h}(\textbf{S}_{i} + \textbf{Z}_{i}^{N}) + \textbf{n}_{B}, \\
			&= \sqrt{1-\sigma^{2}} \textbf{h}_{est} \textbf{S}_{i} \\
			&+ \underbrace{\sqrt{\sigma^{2}} \textbf{h}_{err} \textbf{S}_{i} + \sqrt{\frac{\sigma^2 (1-\alpha) P_{tot}}{P_{Z}}} \textbf{h}_{err} \textbf{Z}_{i} + \textbf{n}_{B}}_{\hat{\textbf{n}}_{B}}
			\label{eq6}
		\end{split}
	\end{align}
	}where $\textbf{Z}_{i}^{N}$ represents the normalized AN matrix, and $P_{Z}$ represents the power of the AN matrix, where $P_{Z} = 8$, $\textbf{n}_{B}$ is the complex additive white Gaussian noise (AWGN) vector with zero-mean and $N_{0}$ variance entries, and $\hat{\textbf{n}}_{B}$ is the colored Gaussian noise vector of which the entries are uncorrelated with the variance of $\hat{N}_{B}=\mathbb{E}[\hat{\textbf{n}}_{B}\hat{\textbf{n}}_{B}^{H}]/4$. Thus, the received signal at Bob can be rewritten by

	{\footnotesize
	\begin{align}
		\begin{split}
			\textbf{y}_{B} = \sqrt{1-\sigma^{2}} \textbf{h}_{est} \textbf{S}_{i} + \hat{\textbf{n}}_{B}.
			\label{eq7}
		\end{split}
	\end{align}
	}Since the entries of $\hat{\textbf{n}}_{B}$ are colored, the linear whitening transformation function $(\mathit{\Psi_{B}}=\sqrt{N_{0}}(\hat{N}_{B})^{-1/2})$ is applied on each entry to make them white Gaussian noise entries with zero-mean and $N_{0}$ variance \cite{8_2019_Conventional}. Thus, the processed received signal at Bob becomes $\tilde{\textbf{y}}_{B}=\tilde{\textbf{h}}_{est} \textbf{S}_{i} + \tilde{\textbf{n}}_{B}$, where $\tilde{\textbf{y}}_{B}=\mathit{\Psi_{B}} \textbf{y}_{B}$, $\tilde{\textbf{h}}_{est}=\mathit{\Psi_{B}} \sqrt{1-\sigma^{2}} \textbf{h}_{est}$, and $\tilde{\textbf{n}}_{B}=\mathit{\Psi_{B}} \hat{\textbf{n}}_{B}$.
	
	The received signal model of Eve is almost the same with Bob, where the only difference is the AN, which is given as
	
	{\footnotesize
	\begin{align}
		\begin{split}
			\textbf{y}_{E} = \textbf{g}(\textbf{S}_{i} + \textbf{Z}_{i}^{N}) + \textbf{n}_{E} 
			= \sqrt{1-\sigma^{2}} \textbf{g}_{est} \textbf{S}_{i} + \hat{\textbf{n}}_{E}
			\label{eq8}
		\end{split}
	\end{align}
	}where $\textbf{n}_{E}$ is the AWGN vector with zero-mean and $N_{0}$ variance elements, and $\hat{\textbf{n}}_{E}$ is the colored Gaussian noise vector of which the entries are uncorrelated with the variance of $\hat{N}_{E}=\mathbb{E}[\hat{\textbf{n}}_{E}\hat{\textbf{n}}_{E}^{H}]/4$, which is given in \eqref{eq9}. Applying the linear whitening transformation function $(\mathit{\Psi_{E}}=\sqrt{N_{0}}(\hat{N}_{E})^{-1/2})$, the received signal at Eve can be obtained as $\tilde{\textbf{y}}_{E}=\tilde{\textbf{g}}_{est} \textbf{S}_{i} + \tilde{\textbf{n}}_{E}$, where $\tilde{\textbf{y}}_{E}=\mathit{\Psi_{E}} \textbf{y}_{E}$, $\tilde{\textbf{g}}_{est}=\mathit{\Psi_{E}} \sqrt{1-\sigma^{2}} \textbf{g}_{est}$, and $\tilde{\textbf{n}}_{E}=\mathit{\Psi_{E}} \hat{\textbf{n}}_{E}$.

	\begin{figure*}
		{\footnotesize
		\begin{align}
			\begin{split}
				\hat{\textbf{n}}_{E} = \sqrt{\frac{(1-\sigma^{2})(1-\alpha)P_{tot}}{P_{Z}}} \textbf{g}_{est} \textbf{Z}_{i} + \sqrt{\sigma^{2}} \textbf{g}_{err} \textbf{S}_{i} + \sqrt{\frac{\sigma^{2}(1-\alpha)P_{tot}}{P_{Z}}} \textbf{g}_{err} \textbf{Z}_{i} + \textbf{n}_{E}.
				\label{eq9}
			\end{split}
		\end{align}
		}%
		\hrulefill
		\vspace*{-4ex}
	\end{figure*}

	\subsubsection{Detection}
	Both Bob and Eve use the two-stage detection method, which is a symbol-by-symbol and low complexity detection scheme. At the first stage, antenna combination is detected via summing minimum decision metrics of each symbol, which is given by
	
	{\footnotesize	
		\begin{align}
			\begin{split}
				\hat{i} &= \argmin_{i \in \{0,\dots,\frac{N}{2}-1\}} \sum\nolimits_{k = 1}^{4} \epsilon_{r}^{i,k}, \quad r \in \{B,E\}
				\label{eq10}
			\end{split}
		\end{align}
	}where $\epsilon_{B}^{i,k} = \|\tilde{\textbf{y}}_{B} - \tilde{\textbf{h}}_{est} (\textbf{A}_{2k-1,i} x_{\zeta_{B}^{i,k} I} + \textbf{A}_{2k,i} x_{\zeta_{B}^{i,k} Q})\|^{2}$ and $\epsilon_{E}^{i,k} = \|\tilde{\textbf{y}}_{E} - \tilde{\textbf{g}}_{est} (\textbf{A}_{2k-1,i} x_{\zeta_{E}^{i,k} I} + \textbf{A}_{2k,i} x_{\zeta_{E}^{i,k} Q})\|^{2}$ represent the metric of the $k^{\mathrm{th}}$ symbol when $i^{\mathrm{th}}$ antenna combination is selected for Bob and Eve, respectively, while $\zeta_{B}^{i,k} = \argmin_{\zeta \in \{1,\dots,M\}} \|\tilde{\textbf{y}}_{B} - \tilde{\textbf{h}}_{est} (\textbf{A}_{2k-1,i} x_{\zeta I} + \textbf{A}_{2k,i} x_{\zeta Q})\|^{2}$ and $\zeta_{E}^{i,k} = \argmin_{\zeta \in \{1,\dots,M\}} \|\tilde{\textbf{y}}_{E} - \tilde{\textbf{g}}_{est} (\textbf{A}_{2k-1,i} x_{\zeta I} + \textbf{A}_{2k,i} x_{\zeta Q})\|^{2}$ represent the symbol index that provides minimum metric of the $k^{\mathrm{th}}$ symbol when $i^{\mathrm{th}}$ antenna combination is selected for Bob and Eve, respectively. At the second stage, each symbol is separately detected as $\hat{x}_{1} = \Omega(\zeta_{r}^{\hat{i},1})$, $\hat{x}_{2} = \Omega(\zeta_{r}^{\hat{i},2})$, $\hat{x}_{3} = \Omega(\zeta_{r}^{\hat{i},3})$, and $\hat{x}_{4} = \Omega(\zeta_{r}^{\hat{i},4})$.
	\vspace{-1ex}		
	\section{Performance Analysis}
	In this section, we derive the ESR and theoretical BER expressions in order to evaluate the secrecy and error performance and verify our computer simulations.
	\vspace{-2ex}		
	\subsection{Ergodic Secrecy Rate}
	There are a total of $N/2$ antenna combinations and each APM symbol is chosen from the normalized $M$-ary constellation. Assuming that $\chi$ is the set of all possible CIOD matrices, the elements of $\chi$ are discrete variables with the same probability of $2/NM^{4}$, and $\textbf{X}_{n}$ denotes the $n^{\mathrm{th}}$ CIOD matrix in the set. The ESR of the CIOD-IM scheme is expressed as $R_{S} = [R_{B} - R_{E}]^{+}$ where $[a]^{+}$ indicates $\max\{0, a\}$, $R_{B}$ and $R_{E}$ indicate the ergodic rate of Bob and Eve, respectively. Since the received signal at Bob can be given as $\tilde{\textbf{y}}_{B} = \tilde{\textbf{h}}_{est} \textbf{X}_{n} + \tilde{\textbf{n}}_{B}$, its complex received vector has the conditional probability density function (PDF) that is given by \cite{12_2012_OnSecrecyMutual}
	
	{\footnotesize
	\begin{align}
		\begin{split}
			p(\tilde{\textbf{y}}_{B} | \textbf{X}=\textbf{X}_{n}) = \bigg[\frac{1}{\pi N_{0}}\bigg]^{4} \exp\bigg(\frac{-\|\tilde{\textbf{y}}_{B} - \tilde{\textbf{h}}_{est} \textbf{X}_{n}\|^{2}}{N_{0}}\bigg).
			\label{eq11}
		\end{split}
	\end{align}
	}The marginal PDF of the complex received vector is expressed as

	{\footnotesize
	\begin{align}
		\begin{split}
			p(\tilde{\textbf{y}}_{B}) = \frac{2}{N M^{4}} \sum\nolimits_{n = 1}^{\frac{N}{2} M^{4}} \bigg[\frac{1}{\pi N_{0}}\bigg]^{4} \exp\bigg(\frac{-\|\tilde{\textbf{y}}_{B} - \tilde{\textbf{h}}_{est} \textbf{X}_{n}\|^{2}}{N_{0}}\bigg).
			\label{eq12}
		\end{split}
	\end{align}
	}The mutual information of Bob can be obtained as follows

	{\footnotesize 
	\begin{align}
		\begin{split}
			&I(\tilde{\textbf{y}}_{B};\textbf{X}) = \sum_{n} \int_{\tilde{\textbf{y}}_{B}} p(\textbf{X},\tilde{\textbf{y}}_{B}) \log_{2} \frac{p(\textbf{X},\tilde{\textbf{y}}_{B})}{p(\textbf{X}) p(\tilde{\textbf{y}}_{B})} d\tilde{\textbf{y}}_{B}, \\
			&= \log_{2} \frac{N}{2} M^{4} - \frac{2}{N M^{4}} \sum\nolimits_{n = 1}^{\frac{N}{2} M^{4}} \mathbb{E}_{\tilde{\textbf{n}}_{B}} \bigg[\log_{2} \sum\nolimits_{n_{2} = 1}^{\frac{N}{2} M^{4}} \\ &\exp\bigg(-\frac{\|\tilde{\textbf{h}}_{est}(\textbf{X}_{n} - \textbf{X}_{n_{2}}) + \tilde{\textbf{n}}_{B}\|^{2} - \|\tilde{\textbf{n}}_{B}\|^{2}}{N_{0}}\bigg)\bigg].
			\label{eq13}
		\end{split}
	\end{align}
	}Since the transmission is conducted in four time slots, the ergodic rate of Bob can be obtained as $R_{B}=I(\tilde{\textbf{y}}_{B};\textbf{X})/4$.
	
	Using a similar approach, the mutual information and the ergodic rate of Eve can be obtained as given
	
	{\footnotesize
	\begin{align}
		\begin{split}
			I(\tilde{\textbf{y}}_{E};\textbf{X}) &= \log_{2} \frac{N}{2} M^{4} - \frac{2}{N M^{4}} \sum_{n = 1}^{\frac{N}{2} M^{4}} \mathbb{E}_{\tilde{\textbf{n}}_{E}} \bigg[\log_{2} \sum_{n_{2} = 1}^{\frac{N}{2} M^{4}} \\ \exp\bigg(&-\frac{\|\tilde{\textbf{g}}_{est}(\textbf{X}_{n} - \textbf{X}_{n_{2}}) + \tilde{\textbf{n}}_{E}\|^{2} - \|\tilde{\textbf{n}}_{E}\|^{2}}{N_{0}}\bigg)\bigg], \\
			R_{E} &= \frac{I(\tilde{\textbf{y}}_{E};\textbf{X})}{4}.
			\label{eq14}
		\end{split}
	\end{align}
	}
	\vspace{-5ex}
	\subsection{Theoretical Bit Error Rate}
	Utilizing the union bound technique with associated pairwise error probabilities (PEPs), the theoretical BER upper bound is derived for the case of perfect channel estimation. Assuming that $\textbf{X}_{u}$ is erroneously detected when $\textbf{X}_{n}$ is transmitted, the theoretical BER upper bound can be given as \cite{13_2016_QSMNakagami}

	{\footnotesize
	\begin{align}
		\begin{split}
			P_{b} \leq \frac{2}{NM^{4}} \sum\nolimits_{n=1}^{\frac{NM^{4}}{2}} \sum\nolimits_{u=1}^{\frac{NM^{4}}{2}} \frac{e_{n,u}}{\log_{2}\big(\frac{NM^{4}}{2}\big)} \bar{P}_{e}(\textbf{X}_{n} \rightarrow \textbf{X}_{u})
			\label{eq15}
		\end{split}
	\end{align}
	}where $\bar{P}_{e}(\textbf{X}_{n} \rightarrow \textbf{X}_{u})$ represents the PEP of deciding $\textbf{X}_{u}$ when $\textbf{X}_{n}$ is transmitted, and $e_{n,u}$ is the number of bit errors for the corresponding pairwise error event. The conditional PEP (CPEP) depending on $\textbf{h}$ can be given by $\bar{P}_{e}(\textbf{X}_{n} \rightarrow \textbf{X}_{u}|\textbf{h}) = Q(\sqrt{\gamma_{S} \gamma})$ where $\gamma_{S} = \alpha P_{tot}/2N_{0}$, and $\gamma=\|\textbf{h}\boldsymbol{\Phi}_{nu}\|^{2}$ in which $\boldsymbol{\Phi}_{nu}=\textbf{X}_{n}-\textbf{X}_{u}$ with $\textbf{X}_{n}, \textbf{X}_{u} \in \chi$. The unconditional PEP can be obtained by taking the expectation of CPEP, which can be represented as $\int_{0}^{\infty} Q(\sqrt{\gamma_{S} \gamma}) p_{\gamma}(\gamma) d\gamma$. Using the moment generating function of $\gamma$, $M_{\gamma}(s)=[\det(\textbf{I}_{4}-s\Delta)]^{-1}=\prod_{d=1}^{D}(1 - s\lambda_{d})^{-1}$ where $\Delta=(\boldsymbol{\Phi}_{nu}^{H} \boldsymbol{\Phi}_{nu})$, $D=\mathrm{rank}(\Delta)$, and $\lambda_{d},d\in\{1,\dots,4\}$ representing the eigenvalues of $\Delta$, and the alternative expression of $Q$-function, the PEP can be obtained as \cite{14_2017_STCM}

	{\footnotesize
	\begin{align}
		\begin{split}
				\bar{P}_{e}(\textbf{X}_{n} \rightarrow \textbf{X}_{u}) = \frac{1}{\pi} \int\nolimits_{0}^{\frac{\pi}{2}} \prod\nolimits_{d=1}^{D}\bigg(1+\frac{\alpha P_{tot}\lambda_{d}}{2N_{0}\sin^{2}\theta}\bigg)^{-1}d\theta.
				\label{eq16}
		\end{split}
	\end{align}
	}
	\vspace{-4ex}
	\section{Simulation Results}
	In this section, BER and ESR performances of the CIOD-IM scheme are investigated with respect to the efficient Alamouti \cite{9_2019_EfficientAlamouti} and the conventional \cite{8_2019_Conventional} schemes, and the BER performances are verified by the theoretical BER upper bound given in \eqref{eq15}. The impact of imperfect channel estimation and power allocation are analyzed in terms of BER and ESR performances, respectively. For the sake of fairness of the comparisons, it is assumed that $P_{tot}=1$ for all schemes. In all simulations, signal-to-noise ratio (SNR) is defined as $E_{s}/N_{0}$, where $E_{s}=\alpha P_{tot}/8$ for the CIOD-IM scheme, $E_{s}=\alpha P_{tot}/4$ for the efficient Alamouti scheme, and $E_{s}=\alpha P_{tot}$ for the conventional scheme, and perfect channel estimation is assumed unless otherwise stated. In addition, the results of the conventional scheme is obtained assuming SLNR based TAS \cite{8_2019_Conventional}. In Fig. 4 and Fig. 6, $N_{a}$ and $N_{t}$ represents the number of total and selected transmit antennas at Alice, respectively. 
	
	Fig. 3 illustrates the BER performance of the CIOD-IM scheme for a large-scale MISO system with $N=4$ and $32$ by employing different modulation levels as QPSK and 16-QAM. It can be seen that as $N$ increases, the BER performance gets worse in the most of the SNR regions while the spectral efficiency rises. Also, we verify our simulation results with theoretical upper bounds for the reference two systems, which shows that our simulation results match with the upper bounds at high SNR values. It can also be observed that Eve's BER is almost 0.5 for all $N$ and $M$ values, which indicates that Eve cannot obtain any information.
	
	In Fig. 4, we compare the BER performance of the CIOD-IM scheme with the efficient Alamouti and the conventional schemes for $2 \leq l \leq 3$. It is clear that the proposed scheme outperforms benchmark schemes. It should be noted that regardless of the number of transmit antennas at Alice, there are only two active RF chains in the CIOD-IM scheme, which ensures improved performance without additional complexity. Diversity orders will be more evident in the higher SNR regimes.
	
	In Figs. 5(a) and 5(b), we investigate the effect of imperfect channel estimation and power allocation on the BER and ESR performances of the CIOD-IM scheme, respectively, for $N=4$ and QPSK. In Fig. 5(a), we considered the cases of $\sigma^{2}=0$ which corresponds to perfect channel estimation, $\sigma^{2}=0.03$, and $\sigma^{2}=0.1$. It is clear that any error in channel estimation significantly deteriorates the BER performance.
	
	Fig. 5(b) illustrates that the security performance degrades as $\alpha$ increases. Also, it can be observed that secrecy behavior is different when most of the power is allocated to APM symbols. As the SNR increases, the more power allocated to APM symbols becomes more significant as channel quality increases. Therefore, secrecy diminishes in the high SNR region.
	
	Finally, the ESR performances of the CIOD-IM and benchmark schemes are compared in Fig. 6. The CIOD-IM scheme achieves a better ESR performance than the efficient Alamouti scheme for the same spectral efficiency in all SNR regions. These results ensure that the CIOD-IM scheme achieves a competitive secrecy rate.\vspace{-1ex}
	
	\begin{figure}[h]
		\vspace{-4.5ex}
		\centering
		\includegraphics[width=53ex]{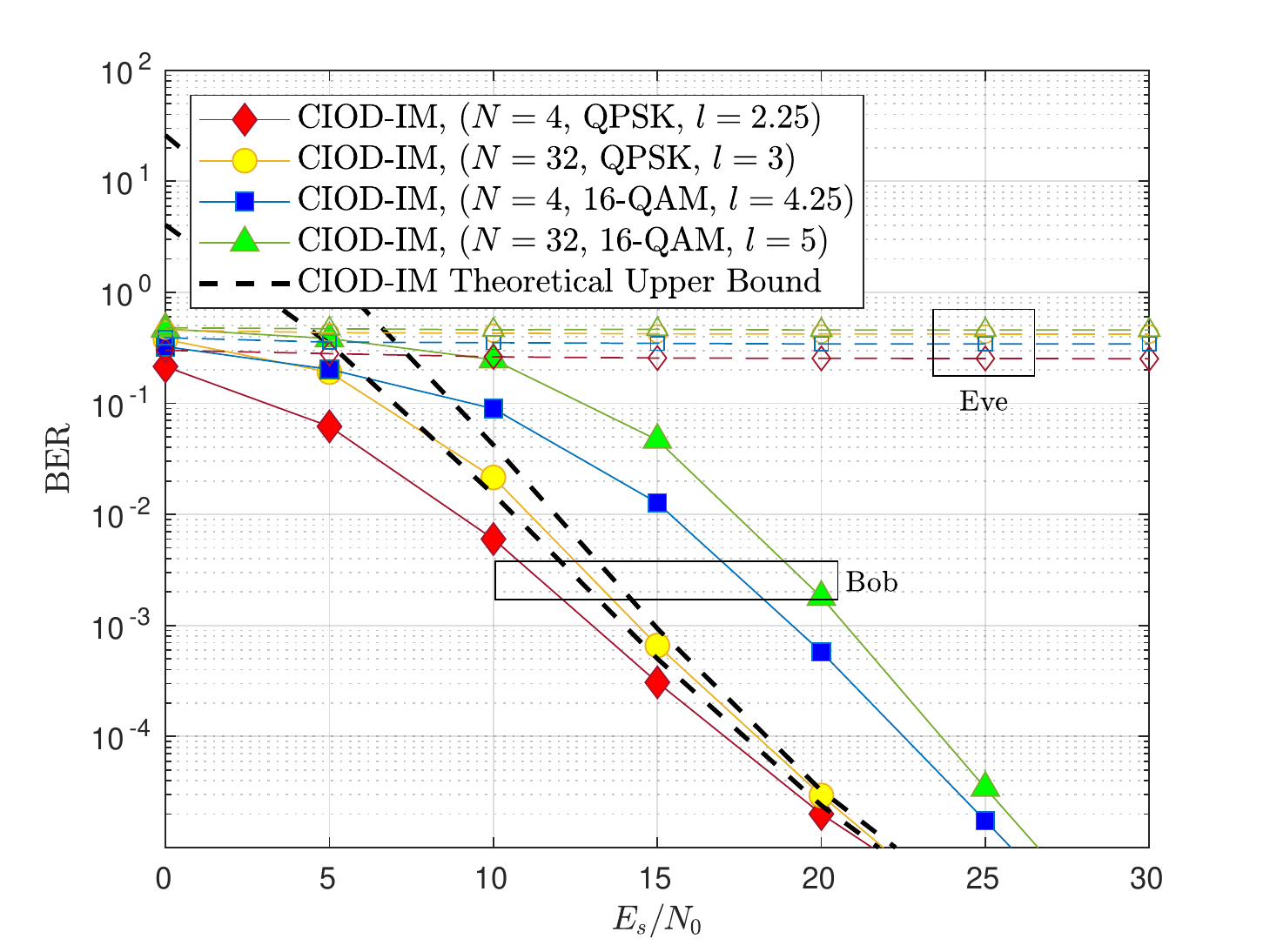}
		\caption{BER performance of the CIOD-IM scheme for $N=4$ and $32$ with QPSK and $16$-QAM.}
		\label{fig:Fig3}
		\vspace{-3ex}
	\end{figure}
	\begin{figure}[h]
		\centering
		\includegraphics[width=53ex]{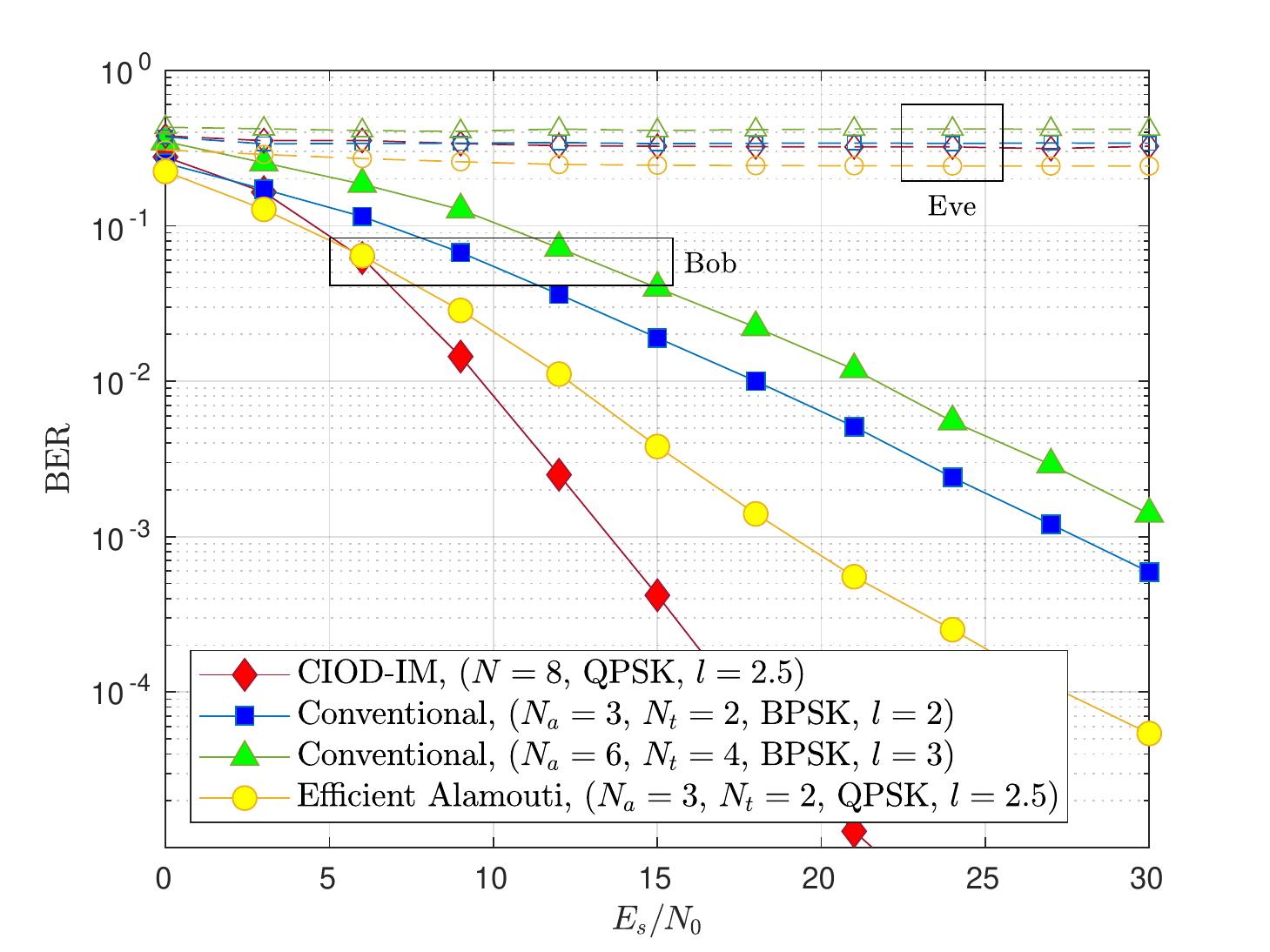}
		\caption{Comparison of the BER performances of the CIOD-IM and benchmark schemes for $2 \leq l \leq 3$.}
		\label{fig:Fig4}
		\vspace{-3.5ex}
	\end{figure} 
	\begin{figure}[h]
		\vspace{-4.5ex}
		\centering
		\includegraphics[width=53ex]{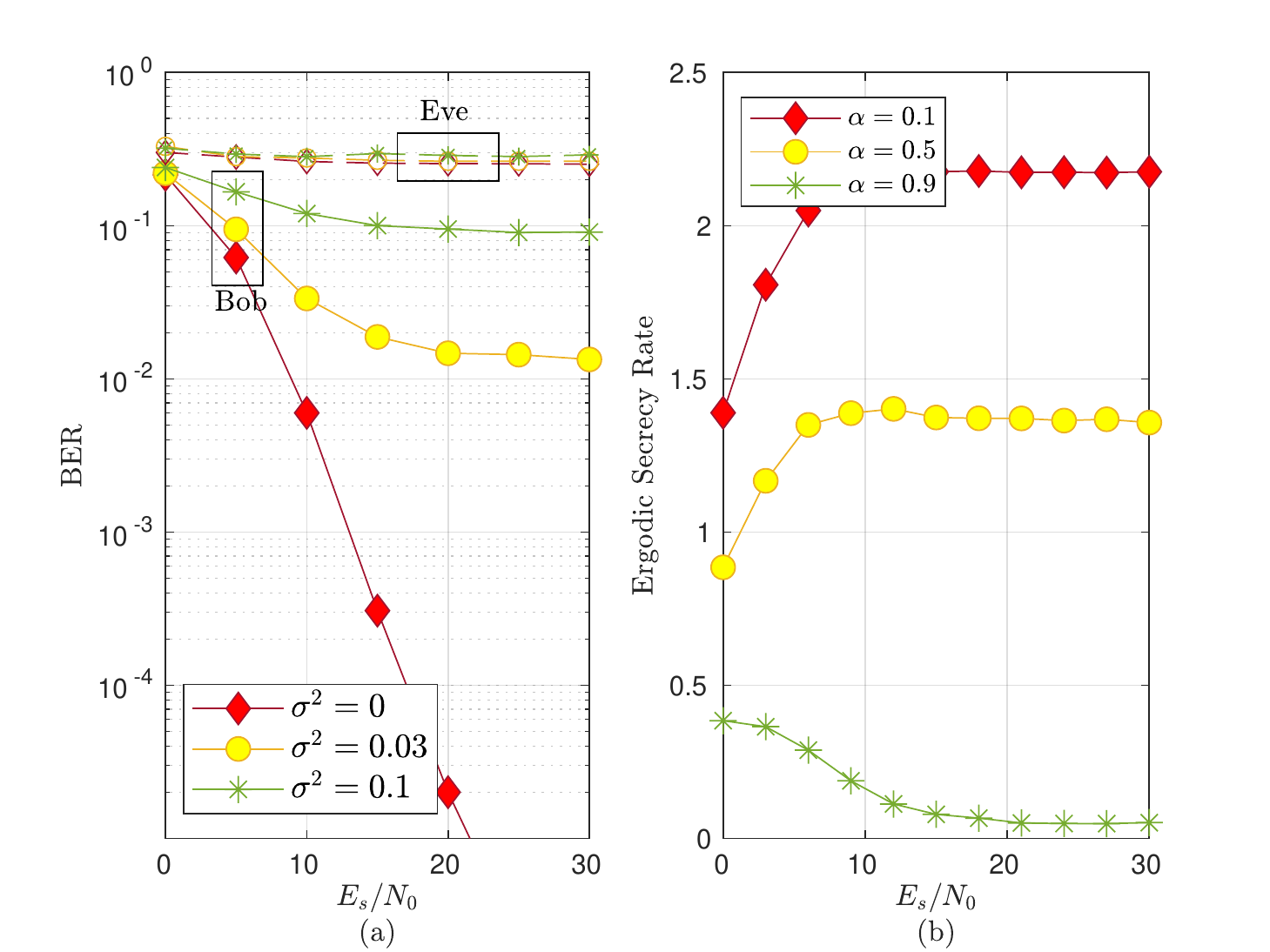}
		\caption{The effect of (a) imperfect CSI on the BER. (b) power allocation on the ESR performance of the CIOD-IM scheme for $N=4$ and QPSK.}
		\label{fig:Fig5}
		\vspace{-3ex}
	\end{figure}
	\begin{figure}[h]
		\centering
		\includegraphics[width=53ex]{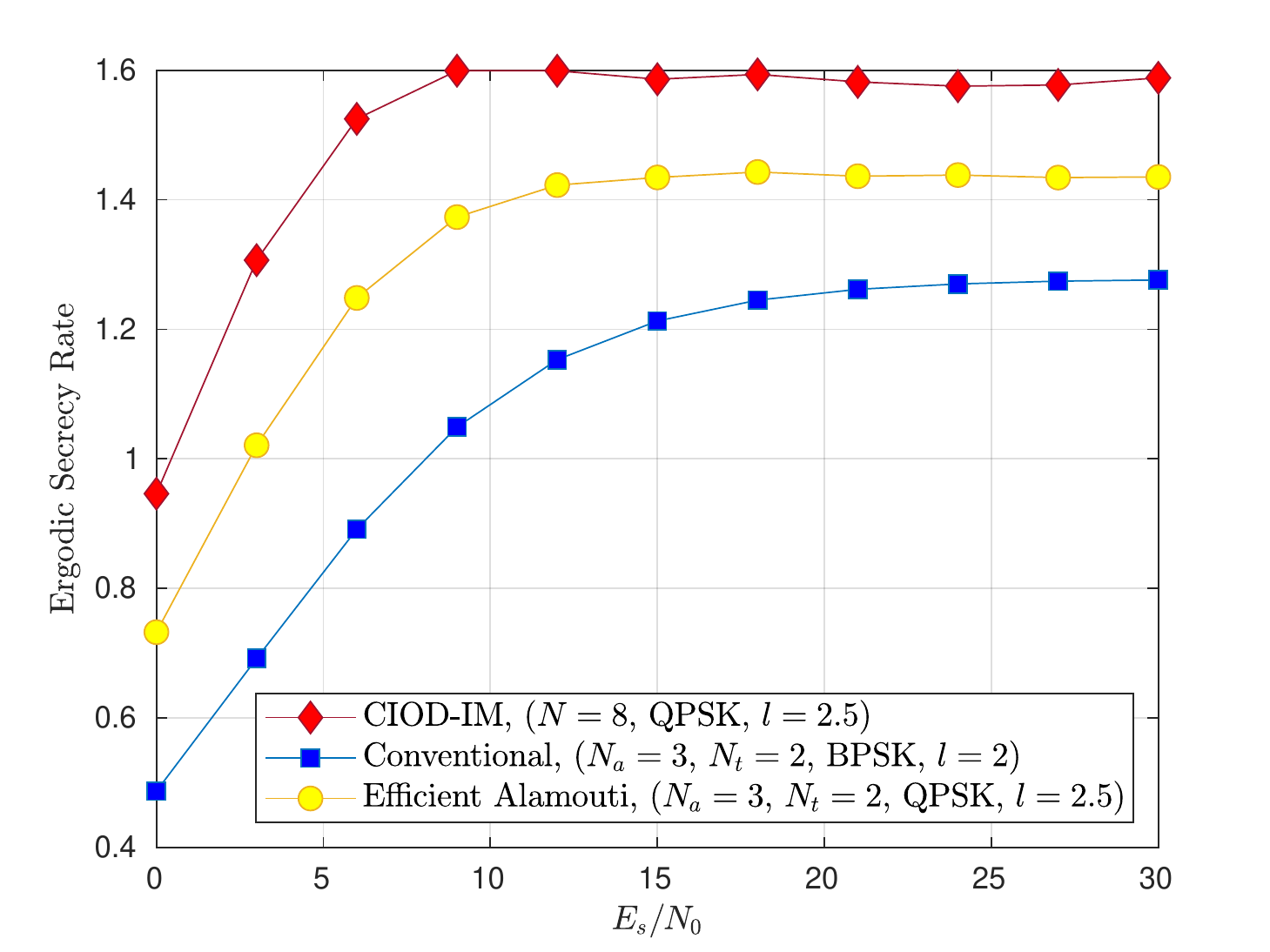}
		\caption{Comparison of the ESR performances of the CIOD-IM and benchmark schemes for $2 \leq l \leq 3$.}
		\label{fig:Fig6}
		\vspace{-3.5ex}
	\end{figure}
	\section{Conclusion}
	\vspace{-1ex}
	In this paper, a new secrecy scheme, which is called as CIOD-IM, has been introduced. The proposed system model increases the spectral efficiency of CIOD transmission by means of a novel IM technique. Moreover, the special design of the AN matrix provides a satisfactory secrecy performance. The security and the BER performances have been investigated under Rayleigh fading environment and compared with benchmark schemes. The ergodic rates of Bob and Eve have been derived and the ergodic secrecy rate has been obtained. In order to verify the correctness of our simulation results, the theoretical upper bound of the BER has been derived. Our simulations show that the CIOD-IM scheme achieves better results than the benchmark schemes both in terms of the secrecy and the BER performances. Finally, it has been observed by its improved BER results in large-scale MISO setups that the CIOD-IM scheme can be a promising candidate for future massive MIMO systems requiring both high reliability and secrecy.\vspace{-2.5ex}
	
	\bibliographystyle{IEEEtran}
	\bibliography{references}

\begin{thebibliography}{10}
\providecommand{\url}[1]{#1}
\csname url@samestyle\endcsname
\providecommand{\newblock}{\relax}
\providecommand{\bibinfo}[2]{#2}
\providecommand{\BIBentrySTDinterwordspacing}{\spaceskip=0pt\relax}
\providecommand{\BIBentryALTinterwordstretchfactor}{4}
\providecommand{\BIBentryALTinterwordspacing}{\spaceskip=\fontdimen2\font plus
\BIBentryALTinterwordstretchfactor\fontdimen3\font minus
  \fontdimen4\font\relax}
\providecommand{\BIBforeignlanguage}[2]{{%
\expandafter\ifx\csname l@#1\endcsname\relax
\typeout{** WARNING: IEEEtran.bst: No hyphenation pattern has been}%
\typeout{** loaded for the language `#1'. Using the pattern for}%
\typeout{** the default language instead.}%
\else
\language=\csname l@#1\endcsname
\fi
#2}}
\providecommand{\BIBdecl}{\relax}
\BIBdecl

\bibitem{1_2017_EBsurvey}
E.~{Basar}, M.~{Wen}, R.~{Mesleh}, M.~{Di Renzo}, Y.~{Xiao}, and H.~{Haas},
  ``Index modulation techniques for next-generation wireless networks,''
  \emph{IEEE Access}, vol.~5, pp. 16\,693--16\,746, 2017.

\bibitem{2_2008_SM}
R.~Y. {Mesleh}, H.~{Haas}, S.~{Sinanovic}, C.~W. {Ahn}, and S.~{Yun}, ``Spatial
  modulation,'' \emph{IEEE Trans. on Veh. Technol.}, vol.~57, no.~4, pp.
  2228--2241, July 2008.

\bibitem{3_2016_Precoding}
F.~{Wu}, R.~{Zhang}, L.~{Yang}, and W.~{Wang}, ``Transmitter precoding-aided
  spatial modulation for secrecy communications,'' \emph{IEEE Trans. on Veh.
  Technol.}, vol.~65, no.~1, pp. 467--471, Jan. 2016.

\bibitem{4_2015_SPSM}
F.~{Wu}, L.~{Yang}, W.~{Wang}, and Z.~{Kong}, ``Secret precoding-aided spatial
  modulation,'' \emph{IEEE Commun. Lett.}, vol.~19, no.~9, pp. 1544--1547,
  Sept. 2015.

\bibitem{5_2016_SecureMUMIMO}
Y.~{Chen}, L.~{Wang}, Z.~{Zhao}, M.~{Ma}, and B.~{Jiao}, ``Secure multiuser
  mimo downlink transmission via precoding-aided spatial modulation,''
  \emph{IEEE Commun. Lett.}, vol.~20, no.~6, pp. 1116--1119, June 2016.

\bibitem{6_2015_SecrecyEnhancement}
L.~{Wang}, S.~{Bashar}, Y.~{Wei}, and R.~{Li}, ``Secrecy enhancement analysis
  against unknown eavesdropping in spatial modulation,'' \emph{IEEE Commun.
  Lett.}, vol.~19, no.~8, pp. 1351--1354, Aug. 2015.

\bibitem{7_2018_ImperfectCSI}
X.~{Yu}, Y.~{Hu}, Q.~{Pan}, X.~{Dang}, N.~{Li}, and M.~H. {Shan}, ``Secrecy
  performance analysis of artificial-noise-aided spatial modulation in the
  presence of imperfect csi,'' \emph{IEEE Access}, vol.~6, pp.
  41\,060--41\,067, 2018.

\bibitem{8_2019_Conventional}
W.~{Yu}, K.~{Zhang}, P.~{Shang}, X.~{Jiang}, M.~{Wen}, J.~{Li}, and H.~{Hai},
  ``Security enhancing spatial modulation using antenna selection and
  artificial noise cancellation,'' in \emph{Proc. 2019 Int. Conf. on Comput.,
  Netw. and Commun. (ICNC)}, 2019, pp. 105--109.

\bibitem{9_2019_EfficientAlamouti}
P.~{Shang}, S.~{Kim}, and X.~{Jiang}, ``Efficient alamouti-coded spatial
  modulation for secrecy enhancing,'' in \emph{Proc. 2019 Int. Conf. on Inf.
  and Commun. Technol. Convergence (ICTC)}, 2019, pp. 860--864.

\bibitem{10_2004_RectangularCIOD}
M.~Z. A.~K. Khan, B.~S. Rajan, and M.~H. Lee, ``Rectangular co-ordinate
  interleaved orthogonal designs,'' in \emph{Proc. 2003 IEEE Global Telecommun.
  Conf. (GLOBECOM)}, San Francisco, CA, USA, Dec. 2003, pp. 2004--2009.

\bibitem{11_2006_OnSTBCCIOD}
D.~N. {Dao} and C.~{Tellambura}, ``On space-time block codes from coordinate
  interleaved orthogonal designs,'' in \emph{Proc. 2006 IEEE Military Commun.
  Conf. (MILCOM)}, 2006, pp. 1--5.

\bibitem{12_2012_OnSecrecyMutual}
X.~{Guan}, Y.~{Cai}, and W.~{Yang}, ``On the secrecy mutual information of
  spatial modulation with finite alphabet,'' in \emph{Proc. 2012 Int. Conf. on
  Wireless Commun. and Signal Process. (WCSP)}, 2012, pp. 1--4.

\bibitem{13_2016_QSMNakagami}
A.~{Younis}, R.~{Mesleh}, and H.~{Haas}, ``Quadrature spatial modulation
  performance over nakagami- $m$ fading channels,'' \emph{IEEE Trans. on Veh.
  Technol.}, vol.~65, no.~12, pp. 10\,227--10\,231, Dec. 2016.

\bibitem{14_2017_STCM}
E.~{Basar} and I.~{Altunbas}, ``Space-time channel modulation,'' \emph{IEEE
  Trans. on Veh. Technol.}, vol.~66, no.~8, pp. 7609--7614, Aug. 2017.

\end{thebibliography}
\end{document}